\documentclass[twocolumn]{aastex631}

\usepackage{inputenc}
\usepackage{color, amssymb, amsmath, natbib, color} 
\usepackage{multirow, placeins, subfigure}
\usepackage{graphicx}
\usepackage{epstopdf}
\usepackage{bm}
\usepackage{overpic}
\usepackage{hyperref}

\newdimen\digitwidth
\setbox1=\hbox{0}
\digitwidth=\wd1
\catcode`"=\active
\def"{\kern\digitwidth}

\def\lzm{\overline{L_z}}

\def\lzs{{\sigma_{L_z}}}
\def\ej{{E_{\mathrm J}}}

\begin{document}

\title{An Empirical Proxy for the Second Integral of Motion\\
in Rotating Barred or Tri-axial Potentials}

\author[0000-0003-3658-6026]{Yu-Jing Qin}
\affiliation{Steward Observatory, University of Arizona, 933 N Cherry Ave, Tucson, AZ 85721, USA}
\author[0000-0001-5604-1643]{Juntai Shen}%
\correspondingauthor{Juntai Shen}
\email{jtshen@sjtu.edu.cn}
\affiliation{Department of Astronomy, School of Physics and Astronomy, Shanghai Jiao Tong University, 800 Dongchuan Road, Shanghai 200240, China}
\affiliation{Key Laboratory for Particle Astrophysics and Cosmology (MOE) / Shanghai Key Laboratory for Particle Physics and Cosmology, Shanghai 200240, China}
\affiliation{Shanghai Astronomical Observatory, Chinese Academy of Sciences, 80 Nandan Road, Shanghai 200030, China}

\date{\today}

\begin{abstract}
We identify an effective proxy for the analytically unknown second integral of motion ($I_2$)
for rotating barred or tri-axial potentials. Planar orbits of a given energy follow a tight sequence in the space of the time-averaged angular momentum and its amplitude of fluctuation. The sequence monotonically traces the main orbital families in the Poincar\'{e} map, even in the presence of resonant and chaotic orbits. This behavior allows us to define the ``Calibrated Angular Momentum,'' the average angular momentum ($\lzm$) normalized by the amplitude of its fluctuation ($\lzs$), as a numerical proxy for $I_2$. It also implies that the amplitude of fluctuation in $L_z$, previously underappreciated, contains valuable information. This new proxy allows one to classify orbital families easily and accurately, even for real orbits in $N$-body simulations of barred galaxies. It is a good diagnostic tool of dynamical systems, and may facilitate the construction of equilibrium models.
\end{abstract}

\keywords{Galaxy Dynamics, Orbits, Stellar Dynamics, Barred Spiral Galaxies}

\section{Introduction}
\label{sec:intro}
An integral of motion (IoM) $I(\mathbf{x}, \mathbf{v})$ is any time-independent function of the phase-space coordinates that is strictly conserved along an orbit. The \textit{isolating} IoMs, unlike the non-isolating ones, are of great importance as they reduce the dimensionality of the phase space non-trivially and impose fundamental constraints on a dynamical system \citep{BT08}.

Isolating IoMs can be categorized into classical IoMs and nonclassical IoMs. Classical IoMs have known analytical expressions, while nonclassical IoMs do not. Examples of classical IoMs include the Hamiltonian in any time-independent potential, all components of the angular momentum vector in spherical potentials, and the axial component of angular momentum ($L_z$) for axisymmetric potentials. They reflect the symmetries and conservation laws of the system via Noether's theorem. Higher degree of symmetry in Keplerian and harmonic potentials allows even more isolating IoMs. For separable potentials in a certain coordinate system, the equation of motion can be decomposed to decoupled motions in each direction, and the classical IoMs may be obtained (e.g., St\"{a}ckel potentials).

nonclassical IoMs, on the contrary, do not have known analytic expressions of phase-space coordinates $\mathbf{x}$ and $\mathbf{v}$. Their existence is inferred by the fact that a numerically-integrated orbit at a given energy is confined to a closed invariant curve \citep[e.g.][]{hen_hei_64} in the Poincar\'{e} map, also known as the ``surfaces of section'' (SoS). Realistic potentials often contain such nonclassical integrals, which are usually dubbed as the second ($I_2$) or the third integral ($I_3$), since they are in addition to the classical integrals like $H$ and/or $L_z$.  In a Hamiltonian system with $n$ degrees of freedom, regular orbits which admit $n$ isolating IoMs appear as closed \footnote{However, invariant curves may become non-closed, splitting into two disconnected line segments, under certain conditions in a rotating frame \citep{Binney85,xia_shen_21}. Thus the defining characteristic of regular orbits is quasi-periodicity.} invariant curves in the Poincar\'{e} map, as they are confined to $n$-D toroidal surfaces (``orbital tori''). Conversely, chaotic orbits have fewer than $n$ isolating IoMs, so they are diffusive in the Poincar\'{e} map. The non-reducible frequency components of an orbit also indicate the number of isolating IoMs, where each fundamental frequency corresponds to an action variable ($J$) in action-angle coordinates \citep{Binney82, Binney84, Laskar93, Valluri98}.

For a steady-state system, the distribution function can be parameterized using only the isolating IoMs \citep{LyndenBell62}. Given their elegant properties and fundamental roles in classical mechanics, IoMs are extensively used as diagnostic tools for dynamical systems. 

Here we focus on the nonclassical IoMs in rotating barred potentials which are common and useful in astronomy. Nearly two-thirds of disk galaxies in the Universe contain a central elongated bar structure \citep[e.g.][]{MenendezDelmestre07, Masters11}, including our own Milky Way \citep[e.g.][]{Blitz91, binney_etal_91}. Bars are not static structures: in a disk galaxy they rotate rapidly. Tri-axial elliptical galaxies may also have some figure rotation similar to barred galaxies. The orbital structure in a rotating barred potential has been extensively studied \citep{Contopoulos89, Sellwood93, Patsis02, Skokos02, BT08}; the common regular orbital families include prograde $x_1$ orbits, retrograde $x_4$ orbits, and $x_2$/$x_3$ orbits if an Inner Lindblad Resonance (ILR) exists.

In a rotating non-axisymmetric potential, the only classical IoM is the Jacobi integral ($H_\mathrm{J}=H-\mathbf{\Omega}\cdot\mathbf{L}$), which is the Hamiltonian in the rotating frame. Note that the angular momentum is not an IoM in either the inertial or bar-corotating frame. The Poincar\'{e} map for 2D planar orbits in a rotating barred potential shows clear nested invariant curves, implying that at least some orbits are confined to orbital tori by an additional IoM ($I_2$) besides $H_\mathrm{J}$.


Our main motivation is to identify an empirical proxy for the analytically-unknown $I_2$ in a rotating barred potential, which can immediately facilitate the classification of orbits of a given energy. Angular momentum is time-varying in a rotating barred potential, but interestingly we can indeed find such a proxy in the angular momentum space. The amplitude of fluctuation in $L_z$ may also contain valuable information on orbits but was not given enough attention in the past. With simple 2D orbits of test particles, we demonstrate that the ``Calibrated Angular Momentum'', the average angular momentum ($\lzm$) normalized by the amplitude of its fluctuation ($\lzs$), is an excellent numerical proxy for $I_2$.

\section{Potential and Orbital Integration}
Without loss of generality, we adopt a rotating logarithmic bar potential of the following form \citep[Equation 3.103 in][]{BT08}:
\begin{equation}
\label{eq:phi}
\Phi_{\mathrm{L}}(x, y)=\frac{1}{2} v_{0}^{2} \ln \left(R_{\mathrm{c}}^{2}+x^{2}+\frac{y^{2}}{q^{2}}\right) \quad(0<q \leq 1).
\end{equation}
$\Phi_{\mathrm{L}}$ is stationary in a frame that rotates at angular speed $\Omega_{\mathrm b}$.

At $R=\sqrt{(x^2+y^2/q^2)} \ll R_{\mathrm{c}}$, $\Phi_{\mathrm{L}}$ approximates the potential of the two-dimensional harmonic oscillator. At $R \gg R_\mathrm{c}$ and $q \simeq 1$, $\Phi_{\mathrm{L}}\simeq v_{0}^{2} \ln R$, which yields a nearly constant circular speed curve as observed in many disk galaxies.

The equipotentials of $\Phi_{\mathrm{L}}$ have constant axial ratio $q$. The axial ratio $q_\rho \equiv b/a$ of the isodensity surfaces at large radius is \citep[Equation 2.72b in][]{BT08}:
\begin{equation}
\label{eq:q}
q_{\rho}^{2} \simeq q^{4}\left(2-\frac{1}{q^{2}}\right) \quad\left(R \gg R_{\mathrm{c}}\right).
\end{equation}

In our standard model, we adopt $R_{\mathrm{c}}=0.1$, $q=0.84$ (i.e., $q_\rho \simeq 0.54$ according to Equation~\ref{eq:q}), $v_0=1$, and $\Omega_{\mathrm b} = 1$. The units of length, velocity, and acceleration are arbitrary. These parameters place the bar Corotation Radius (CR) at $R_\mathrm{CR}=0.995$ (also the position for the $\mathrm{L}_{1}$ and $\mathrm{L}_{2}$ Lagrangian points). The Jacobi energy at $R_\mathrm{CR}$ is $E_{\mathrm{J,CR}}=-0.495$.

We also tested other analytical and self-consistent $N$-body bar potentials and verified that our main conclusions remain unchanged.

For clarity and cleanness, in this paper we focus mainly on 2D planar orbits of test particles which are numerically integrated. The initial conditions of our test particles are randomly generated inside the equipotential surface to sample all possible orbital families, and the timestep is adaptively adjusted so that each orbit is integrated for about 800 azimuthal periods and each period is sampled by at least 512 points using a 4th-order Runge-Kutta integrator.  The conservation of Jacobi energy $\ej$  is generally better than $4\times 10^{-10}$ for 800 periods. 

\begin{figure*}[!htbp]
\centering
\includegraphics[width=\linewidth]{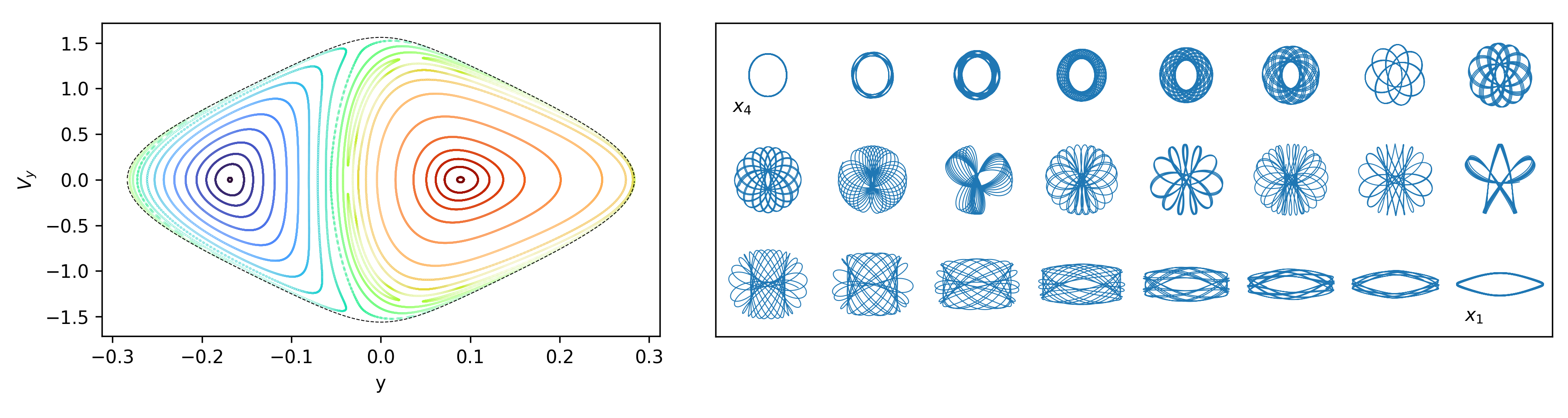}
\caption{
\textit{Left}: (a). Poincar\'{e} SoS at $\ej=-1.08$ with the outmost dashed curve as the zero-velocity curve, i.e., the boundary of the energetically allowed region. Each closed curve corresponds to an orbit. \textit{Right}: (b). The morphology transformation of the corresponding orbits in (a). The first $x_4$ orbit corresponds to the center of left island in (a), then the orbits gradually transition to the last $x_1$ orbit corresponding to the center of right island in (a). The invariant curves in (a) are also  color-coded with the CAM $\equiv \lzm / \lzs = -\cot \phi$ of the orbits. There is a continuous and smooth morphological transition from the periodic $x_4$ orbit to the periodic $x_1$ orbit, following the monotonic increase in CAM (or $\phi$).
\label{fig:sos}}
\end{figure*}

\section{Results}
Figure~\ref{fig:sos}a shows the Poincar\'{e} SoS at $\ej=-1.08$ for our rotating barred potential. Each curve corresponds to an orbit with this $\ej$, and is the record of ($y$, $v_y$) whenever the orbit crosses the bar minor axis ($y$-axis) with $v_x<0$ (to eliminate the sign ambiguity).  Figure~\ref{fig:sos}a shows two predominant regular orbital families within $R_{\mathrm{CR}}$, namely the retrograde $x_4$ family (the left island of nested curves) and the prograde $x_1$ family (the right island). The center points of the two islands are the periodic $x_4$ and $x_1$ orbits, respectively. As discussed in the \S\ref{sec:intro}, the fact that an orbit at a given energy is confined to a closed ``invariant curve'' indicates that this ``regular'' orbit admits an additional nonclassical IoM ($I_2$) at the given $\ej$. Thus, one may regard the invariant curves of regular orbits in the SoS as a contour plot of $I_2$, where $I_2$ changes monotonically as regular orbits transition from one family to another.

Figure~\ref{fig:sos}b shows the morphological transformation of the corresponding orbits in (a). The first orbit is a nearly periodic $x_4$ orbit corresponding to the center of the left island in Figure~\ref{fig:sos}a.  As we move away from the periodic $x_4$ orbit, the enclosed area in the $x_4$ island in the SoS expands, the amplitude of radial oscillations increases, and orbits become thicker rosettes. When we leave the outskirt of $x_4$ island and move into the $x_1$ orbital family, orbits become more ``box-shaped''.  Further along the sequence, orbits become more elongated in the direction of the bar major axis, their enclosed area in SoS gradually shrinks to zero when reaching the periodic $x_1$ orbit, which is the last orbit in Figure~\ref{fig:sos}b corresponding to the center of the right island in Figure~\ref{fig:sos}a. There is a continuous transition of orbital morphology, covering the entire phase space at the given $\ej$, from $x_4$ to $x_1$ families which is accompanied by a monotonic change in $I_2$ respected by each regular orbit.

Searching for a proxy of $I_2$, we study orbits in the space of angular momentum ($L_z$) computed in the co-rotating reference frame of the bar. As expected, $L_z$ is not a conserved quantity and is time-varying. Averaged over time, we can compute the mean angular momentum ($\lzm$) and the standard deviation ($\lzs$) of an orbit, where  
\[ \lzs\equiv\sqrt{\overline{[L_z(t) - \lzm]^2}}\]
reflects the amplitude of fluctuation over the duration of orbital integration.
We have tested that halving or doubling the total time duration of orbital integration changes the values of $\lzm$ and $\lzs$ by only $\la 1\%$ for regular orbits, but they do change by a larger fraction for chaotic orbits (see discussions regarding Figure~\ref{fig:with_x2}).

\begin{figure}
\includegraphics[width=\linewidth]{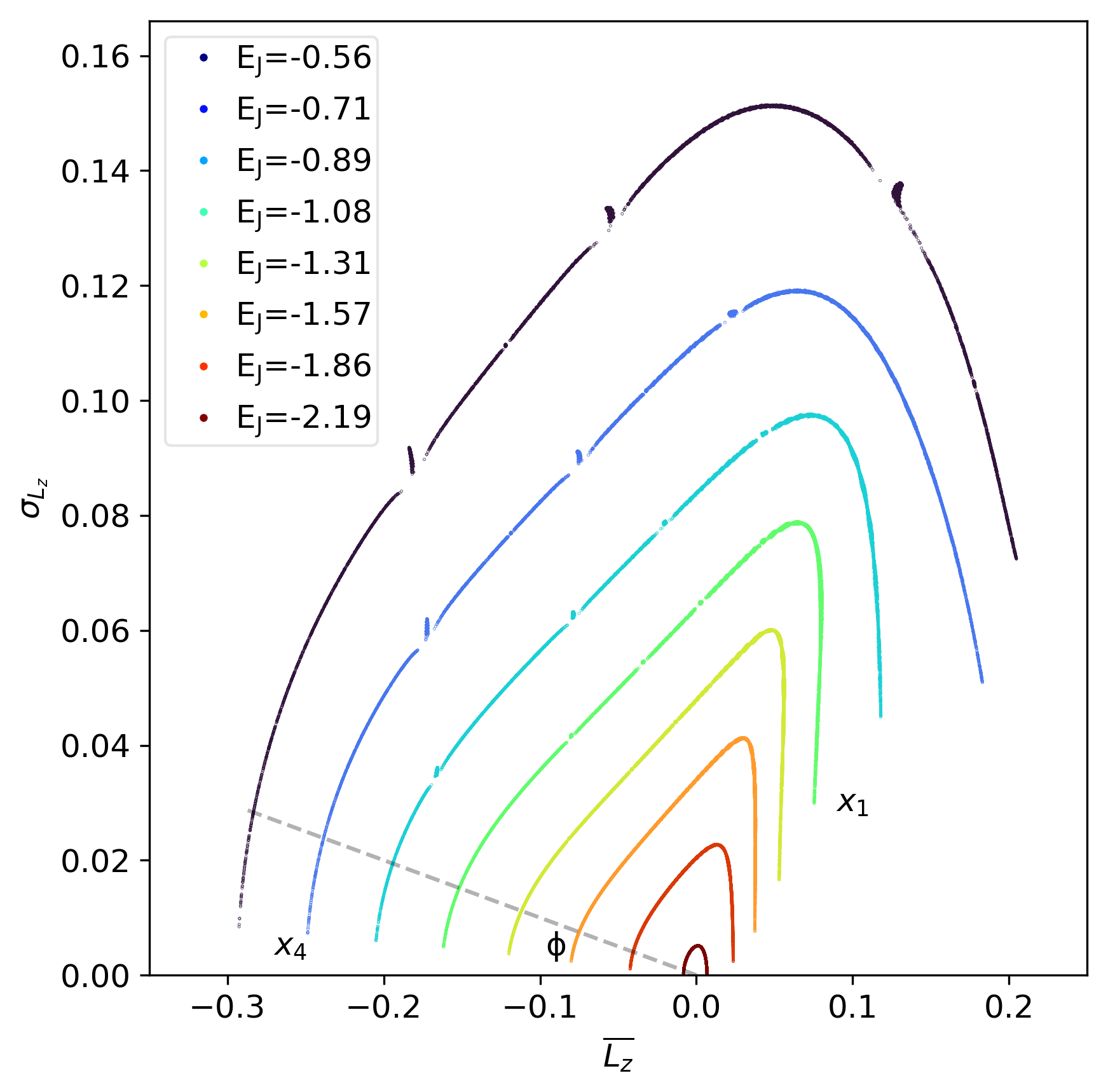}
\caption{Orbital distribution in the angular momentum space ($\lzm$, $\lzs$) for our rotating barred potential. Colors indicate their energies ($\ej$). Note that the green curve ($\ej=-1.08$) corresponds to the orbits shown in Figure~\ref{fig:sos}. An orbital sequence at a given $\ej$ always starts from the periodic $x_4$ (the leftmost point), and ends at the periodic $x_1$ (the rightmost point). }
\label{fig:am_sequence}
\end{figure}

Mapped into the angular momentum space of ($\lzm$, $\lzs$), the orbits with a given $\ej$ follow a compact and nearly continuous sequence, and sequences of different energies are clearly separated in a nested layout (Figure~\ref{fig:am_sequence}). Even more strikingly, the sequence in Figure~\ref{fig:am_sequence} directly and monotonically traces the nested invariant curves in the SoS and the continuous morphological transformation from $x_4$ to $x_1$ orbits. Each sequence in the angular momentum space starts with a periodic $x_4$ orbit at the leftmost end and terminates at a periodic $x_1$ orbit at the rightmost end. Regular orbits connecting the periodic $x_4$ and the periodic $x_1$ in the SoS as illustrated in Figure~\ref{fig:sos} also keep their order in the angular momentum sequence. Such a monotonic mapping from the angular momentum sequence to invariant curves in the SoS indicates that these sequences do trace the monotonic change in $I_2$, hence it can serve as a proxy of $I_2$. The nested compact sequences also imply that both $\lzm$ and $\lzs$ are nearly continuous functions of $\ej$ and $I_2$.

We note that the location in a sequence in Figure~\ref{fig:am_sequence}, which follows $I_2$, can actually be uniquely and monotonically represented by the angle $\phi$ labeled in Figure~\ref{fig:am_sequence}. $\phi$ is equivalent to the Calibrated Angular Momentum (CAM) 
$$\mathrm{CAM} \equiv \lzm/\lzs$$ 
since $\cot \phi=-\lzm/\lzs$. The invariant curves in the SoS (Figure 1a) are also color-coded with CAM. The smooth color variation from $x_4$ to $x_1$ families in the SoS again confirms that CAM indeed monotonically traces $I_2$ for regular orbits. 
Previous works used the average angular momentum to approximate $I_2$ in the vicinity of parent $x_1$/$x_4$ orbits \citep{BT08, Valluri16}. However, the monotonicity of $\lzm$ is not ensured; $\lzm$ increases in the $x_4$ branch, but may decrease slightly in the $x_1$ branch. We have verified that the non-monotonicity of $\lzm$ is more pronounced in bars formed self-consistently from disk instabilities, such as the $N$-body bar model in \citet{Shen10}. Intriguingly, $\lzm$, after being normalized with $\lzs$, then becomes a \textit{monotonic and unique} tracer of the entire phase space at a given $\ej$, smoothly connecting the $x_1$ and $x_4$ families into a single sequence.

\begin{figure}
\includegraphics[width=\linewidth]{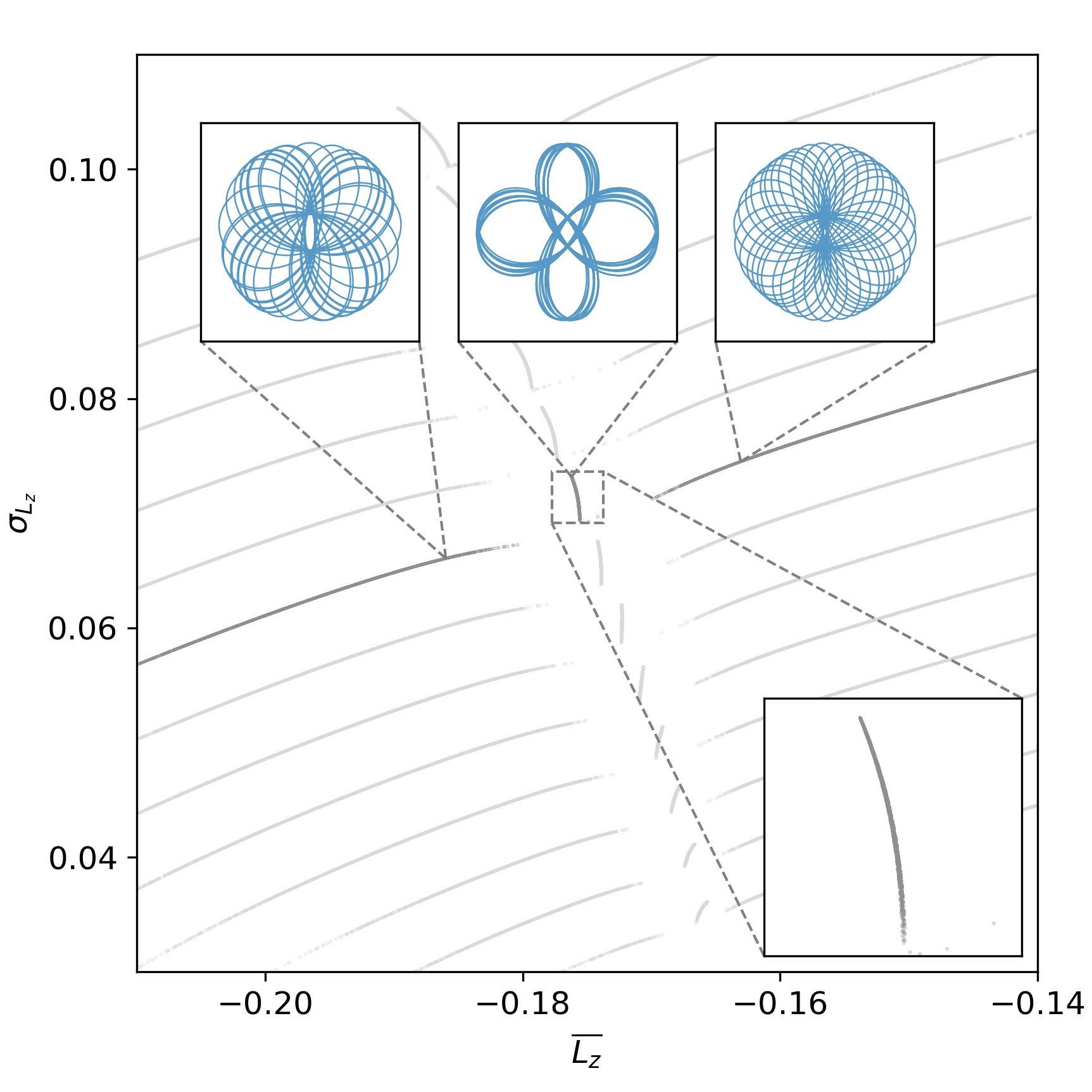}
\caption{
Zoom-in view of ``knots'' in the CAM sequence in Figure~\ref{fig:am_sequence}. The ``knot'' is produced by a 3:4 (azimuthal:radial) resonance (top middle inset). The weak chaos surrounding this resonance creates a discontinuity in the smooth distribution of $\lzm / \lzs$ across a wide range of energies (black points are for $\ej=-0.652$). The top left and top right insets show a smooth transition of orbital morphology near this resonance. 
\label{fig:zoom_in}}
\end{figure}

Figure~\ref{fig:am_sequence} also shows that the sequences in the angular momentum space have tiny breaks and ``knots'' for sufficiently high $\ej$. Figure~\ref{fig:zoom_in} zooms in on one such break and ``knot'' region with much longer orbital integration (around 6400 periods), and reveals that these features are associated with a high-order resonance. It is well-known that stable high-order resonances can alter their local phase space structure and induce chaos around them. Orbital morphology in the insets of Figure~\ref{fig:zoom_in} clearly shows that the ``knot'' is actually a line segment due to a 3:4 (azimuthal:radial) resonance (lower inset of Figure~\ref{fig:zoom_in}), and the break around the ``knot'' may be related to the weakly chaotic orbits surrounding the resonance.  The top tip of the line segment is a periodic 3:4 resonant orbit. Breaks and ``knots'' of a certain resonance are clearly aligned across different energies, carving out a valley across multiple sequences.
Although high-order resonances and weak chaos appear as small localized breaks in the CAM sequence, they do not affect its global trend. In other words, the continuous sequence in the angular momentum space, also in CAM as the numeric proxy for $I_2$, is \textit{insensitive} to local, microscopic phase space structures like high-order resonances. 

The phase space structure illustrated in Figure~\ref{fig:sos} is almost completely dominated by regular orbits of $x_1$ and $x_4$ families, but a rotating barred potential may also have other major resonant families like $x_2$/$x_3$ and strong chaos around them. Thus, one may wonder if the CAM parameter is still an effective proxy of $I_2$ for regular orbits in the presence of strong chaos.

To study the phase space and the angular momentum space when $x_2$/$x_3$ families and strong chaos are present we reduce the pattern speed to $\Omega_\mathrm{b}=0.4$. Chaotic orbits have only one IoM and are not confined to orbital tori; they eventually fill the phase-space volume bounded by the nearby regular  orbits of the same energy. Thus chaotic orbits will drift in the angular momentum space; their $\lzm$ and $\lzs$ could change significantly after a long period of time. We may trace chaotic orbits using their drift in the angular momentum space.  We cut the orbit into two equal halves and estimate their $(\lzm, \lzs)$ separately. The differences between the two halves ($\Delta \lzm$ and $\Delta \lzs$) may be used to define the ``normalized drift'':
\[\delta l = \sqrt{(\Delta \lzm)^2 + (\Delta \lzs)^2} / \sqrt{\lzm^2 + \lzs^2}. \]
The distribution of $\delta l$ has a sharp break around 0.0224. Thus, we consider those orbits with $\delta l > 0.0224$  as possible chaotic orbits (painted grey in  Figure~\ref{fig:with_x2}). Note that $\delta l$ is also reflected in the narrowness of the angular momentum sequence and the chaotic zone.

Figure~\ref{fig:with_x2} shows the $\lzm$--$\lzs$ distribution of the orbits in the model with $\Omega_\mathrm{b}=0.4$.
Short branches of the $x_2$ orbital family are clearly visible in the lower right corner of Figure~\ref{fig:with_x2}. Similar to the sequences of $x_1$ and $x_4$ families, periodic $x_2$ orbits are at the rightmost ends of these $x_2$ branches, and an $x_2$ orbit moves leftward along the sequence as it gradually deviates from the parent periodic $x_2$ orbit. The presence of an $x_2$ branch also induces a chaotic zone in the angular momentum space.  Compared to the canonical case in Figure~\ref{fig:am_sequence}, the region near the maximum of $\lzs$ on each sequence becomes diffuse, twisted, and dominated by strongly chaotic orbits.

\begin{figure}
\includegraphics[width=\linewidth]{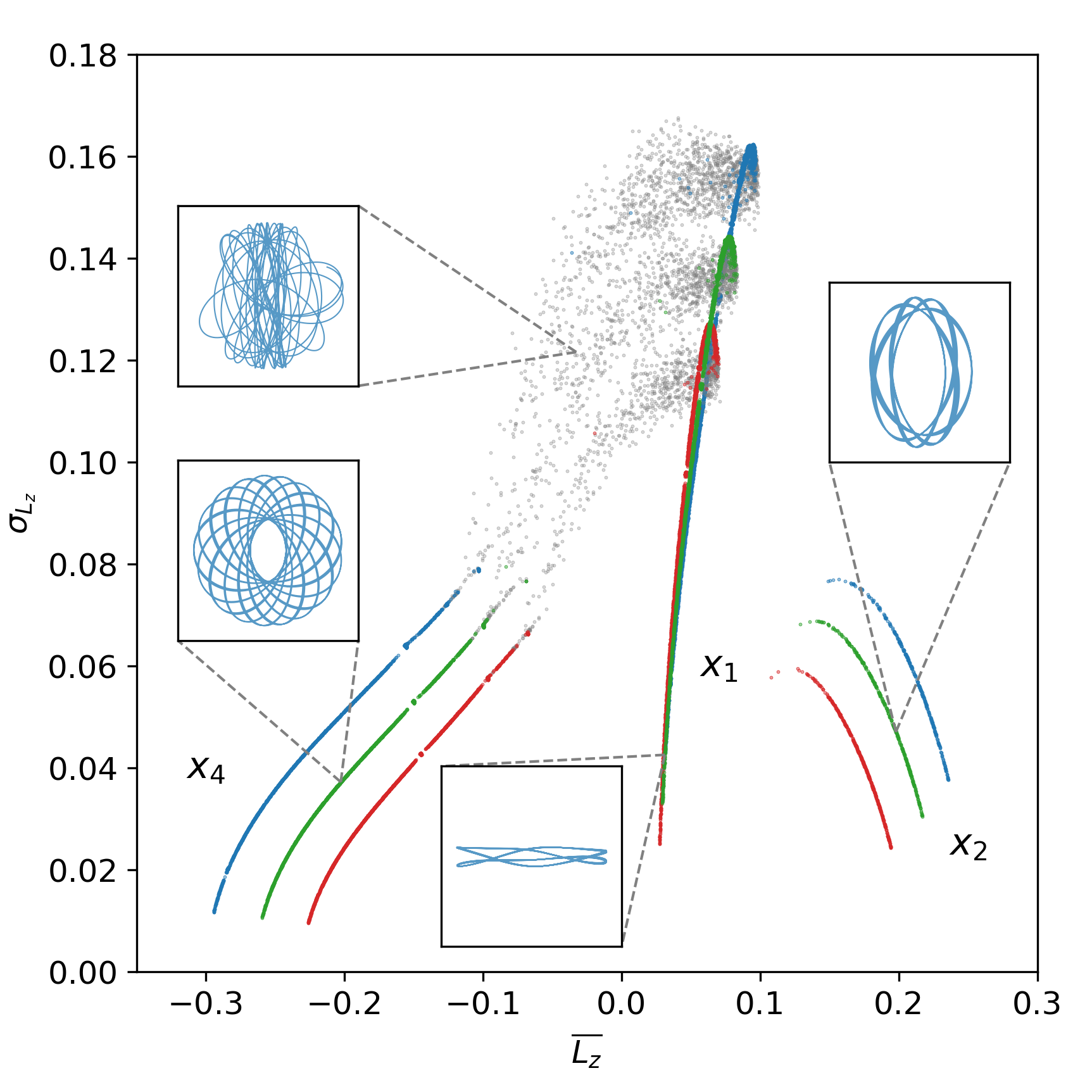}
\caption{Phase space structure for the model with $x_2$ orbits for $\ej=$-0.57, -0.68, and -0.80 (blue, green, red, respectively). Note that the potential's long axis is horizontal. Compared to the canonical case shown in Figure~\ref{fig:am_sequence}, there is a strong, prograde (positive $\lzm$) $x_2$ sub-sequence. There is no sub-sequence for $x_3$ orbital family since it is unstable. Between the sequences of $x_1$ and $x_4$ orbits, there is a diffuse zone of strongly chaotic orbits (grey colored for those with $\delta l>0.0224$). The orbits are integrated for 3200 azimuthal periods. Typical orbits from each regular orbital family and the chaotic zone are illustrated in insets.  
\label{fig:with_x2}}
\end{figure}

The presence of the $x_2$ branch has complicated the phase space structure. Unlike high-order resonances which only alter the local, microscopic phase space structure, the presence of a strong resonance like $x_2$ can significantly reshape the phase space. Although parts of the $x_1$--$x_4$--$x_2$ sequence are interrupted by the presence of strong chaos,  $\lzm/\lzs$ remains valid in tracing the local phase space structure of all stable  islands of the $x_1$, $x_4$, and $x_2$ families. 
Since the angular momentum space is presumably a projection of the $\ej$--$I_2$ space, it is not surprising that chaotic orbits, which do not respect $I_2$, are ``detached'' from the sequences of regular orbits.  Instead, they become ``clouds'' of scattered points in the angular momentum space (grey points in  Figure~\ref{fig:with_x2}). Weakly chaotic orbits diffuse slower, thus they may still remain close to the angular momentum sequence. Surprisingly and fortunately, even chaotic orbits roughly follow the CAM sequence and do not affect much its monotonicity, except for the strongly chaotic zone at the upper tip of the $x_1$ branch, where the monotonicity fails locally. 

\section{Discussions}

\subsection{Asymptotic behaviors in limiting cases}
CAM remains a well-behaved proxy of $I_2$ in most limiting cases as we discuss below. A non-rotating bar, like stationary triaxial potentials, is dominated by box orbits and loop orbits \citep{BT08}. Box orbits (like Lissajous figures) have zero net angular momentum while loop orbits have non-zero net angular momentum. As $\Omega_\mathrm{b}$ gradually approaches zero, $x_4$/$x_2$ orbits in rotating barred potentials become retrograde/prograde loop orbits, and $x_1$ orbits become box orbits with $\lzm=0$ \citep{Valluri16}. 
For the loop orbits, $|\lzm|$ and $\lzs$ form an anti-correlated sequence at a given energy (due to the trade-off between rotation and random motions), thus $\lzm$ is the most obvious proxy of $I_2$. However, CAM, in addition to $\lzm$, still remains a good proxy of $I_2$.
Also, the interconnection of boxy/loop orbits to $x_1$/$x_4$/$x_2$ orbits again reveals the continuity of orbital families in the phase space \citep{Binney84}.

As $q$ in Equation~1 approaches unity, we get an axisymmetric disk which supports only loop orbits with the canonical momentum $p_\phi=L_z$ being an exact integral of motion and $\lzs=0$.
When the axisymmetric disk is viewed in a rotating frame with angular speed $\Omega_\mathrm{b}$, the angular momentum in the corotating frame, $L_z=p_\phi-\Omega_\mathrm{b} R^2$, is no longer an integral. The CAM sequence of loop orbits (corresponding to $x_4$ and $x_1$ in a rotating barred potential) becomes a distorted arch in angular momentum space; the prograde side of the arch is distorted towards $\lzm=0$ and vice versa for the retrograde side. This is similar to the case in Figure~\ref{fig:am_sequence}, where the $x_1$ branch is closer to $\lzm=0$ than the $x_4$ branch. Again, CAM, in addition to $\lzm$, is still a good proxy of $I_2$ in the axisymmetric case despite the fact that $L_z$ (unlike $p_\phi$) is not an integral.

\subsection{Advantages and Potential Applications.}
CAM is a good proxy of $I_2$ even in the presence of high-order resonances and chaotic orbits.  CAM fixes the issue of non-monotonicity of $\lzm$ in tracing the orbital families by taking into account the amplitude of fluctuation in the time-varying $L_z$ ($\lzs$), which contains valuable information but was previously underappreciated. CAM is also independent of the Hamiltonian of the system ($\ej$), and can serve as a good diagnostic of dynamical systems.

We have verified that it is well-behaved in generic rotating bar or tri-axial potentials, including a self-consistent $N$-body bar model designed to match the Milky Way boxy bar/bulge \citep{Shen10}.
It may be generalized to other Hamiltonian systems -- any Hamiltonian that has position-like and momentum-like variables, from which we could construct angular-momentum-like variables. 

There are many potential applications of the empirical proxy. An immediate application of CAM is accurate and quick orbital classification in a barred potential without knowing the detailed properties of the orbital families. This is particularly useful for 3D orbits whose phase space cannot be easily visualized. The CAM orbital classification method is complementary to frequency-based classification methods \citep{Binney82, Valluri98}. Note that neither irreducible frequency components nor frequency ratios are proper substitutions of IoMs.
As a proxy for $I_2$, CAM is more directly related to the fundamentals of orbits. There are less degeneracies associated with CAM ($I_2$) than frequency-based methods since orbits with the same frequency ratio can be distinguished by their intrinsic differences in $\ej$ and $I_2$.  For example, one can easily distinguish $x_1$ and $x_2$ orbits with CAM whereas additional constraints are required to separate them in frequency-based methods (e.g. \citealt{Valluri98, Valluri16}). The computation of CAM is also computationally less expensive than the  Numerical
Analysis of Fundamental Frequencies (NAFF) method \citep{Laskar93}. We have successfully applied it to classify real orbits in $N$-body simulations of barred galaxies, and the results will be presented in a follow-up paper. 

\subsection{Limitations} 
CAM is only an empirical proxy of the analytically-unknown $I_2$. Numerical integration over a sufficient number of periods with \textit{a priori} knowledge of the potential is required to accurately compute CAM. In contrast, a genuine IoM is only a function of phase-space coordinates at \textit{any} moment. It is also important to keep in mind that CAM may not be the only proxy of $I_2$ for a given barred potential.
Several other methods have been developed to estimate or approximate IoMs, particularly action variables, for numerically-integrated orbits in generic triaxial or axisymmetric potentials \citep{Sanders_Binney_14, Sanders_Binney_16}.

\section{Summary}
We discover a good proxy for the second integral of motion $I_2$ in rotating barred or tri-axial models, namely CAM $\equiv \lzm/\lzs$, which monotonically traces various orbital families at a given energy in a rotating barred potential even in the presence of resonant and chaotic orbits.
This empirical proxy of $I_2$ may be used to parameterize pseudo distribution functions (DFs) in the construction of dynamical models, such as Schwarzschild \citep{schwar_79} or made-to-measure models \citep{Syer96}, for real-world observations of rotating barred or triaxial galaxies.
However, we need \textit{a priori} information about the potential to carry out the calculation. Also, CAM-based pseudo DFs do not measure the true phase space density, as CAM is only a dimensionless numerical proxy.

Despite its effectiveness, we still do not fully understand why CAM is such a good proxy of $I_2$. Further study is needed to better understand this relationship. Investigation in the framework of Hamiltonian perturbation theory, and focus on more complicated orbital cases (e.g. real 3D orbits in an $N$-body simulation), could be illuminating.

\begin{acknowledgements}
We thank Scott Tremaine, Jerry Sellwood, and anonymous referee for helpful comments. The research presented here is partially supported by the National Key R\&D Program of China under grant No. 2018YFA0404501; by the National Natural Science Foundation of China under grant Nos. 12025302, 11773052, 11761131016; by the ``111'' Project of the Ministry of Education of China under grant No. B20019; and by the Chinese Space Station Telescope project. J.S. acknowledges support from a {\it Newton Advanced Fellowship} awarded by the Royal Society and the Newton Fund.  This work made use of the Gravity Supercomputer at the Department of Astronomy, Shanghai Jiao Tong University, and the facilities of the Center for High Performance Computing at Shanghai Astronomical Observatory.
\end{acknowledgements}

\bibliography{ref}

\end{document}